\documentclass[twocolumn,showpacs,pra]{revtex4}%
\usepackage{amsfonts}
\usepackage{amsmath}
\usepackage{amssymb}
\usepackage{graphicx}
\usepackage[dvips]{color}%
\providecommand{\U}[1]{\protect\rule{.1in}{.1in}}

\begin{document}
\title{Measurement back-action on the quantum spin-mixing dynamics of a spin-1
Bose-Einstein condensate}
\author{Keye Zhang$^{1}$}
\author{Lu Zhou$^{1}$}
\author{Hong Y. Ling$^{2}$}
\author{Han Pu$^{3}$}
\author{Weiping Zhang$^{1}$}

\affiliation{%
\begin{tabular}
[c]{c}%
$^{1}$State Key Laboratory of Precision Spectroscopy, Dept. of Physics, East
China Normal University, Shanghai 200062, China\\
$^{2}$Department of Physics and Astronomy, Rowan University, Glassboro, New
Jersey 08028-1700, USA\\
$^{3}$Department of Physics and Astronomy, and Rice Quantum Institute, Rice
University, Houston, Texas 77251-1892, USA
\end{tabular}
}

\pacs{03.75.Mn, 03.75.Kk, 03.65.Ta, 42.50.Pq}

\begin{abstract}
We consider a small $F=1$ spinor condensate inside an optical cavity driven by
an optical probe field, and subject the output of the probe to a homodyne
detection, with the goal of investigating the effect of measurement
back-action on the spin dynamics of the condensate. Using the stochastic
master equation approach, we show that the effect of back-action is sensitive
to not only the measurement strength but also the quantum fluctuation of the
spinor condensate. The same method is also used to estimate the atom numbers
below which the effect of back-action becomes so prominent that extracting spin
dynamics from this cavity-based detection scheme is no longer practical.

\end{abstract}
\maketitle

\section{Introduction}

The ability of optical dipole traps to simultaneously cool and trap
ground-state atoms in different magnetic sublevels paved the way for the
experimental realization of spinor Bose-Einstein condensates (BEC)
\cite{Kurn98} where the liberation of spin degrees of freedom has added new
and exciting possibilities in the study of BEC \cite{Ho98}. Of particular
relevance to the present work has been the extensive study of spin dynamics
both in theory \cite{Law98, Kron05, Wenxian05, Heinze10} and in experiments
\cite{Chang04, Schm04, Kron06, Lett07, Lett09}, which aims to understand how
condensate populations exchange coherently among different internal spin
states as well as to explore the potential of these spin oscillations as
probes to the intriguing physics underlying spin-dependent collisions. So far,
such studies have been carried out mainly in spinor condensates with
sufficiently large numbers of atoms where the measurement based on the
standard absorption-imaging technique has established that the spin dynamics
agrees well with the result predicted \cite{Heinze10} within the framework of
mean-field theory \cite{Wenxian05, Kron05}.

The use of the spinor condensates with relatively large atom numbers,
unfortunately, renders it virtually impossible to observe beyond-mean-field
quantum effects, which, besides being fascinating by their own rights, are
thought to be responsible for exotic physics in highly correlated systems. In
small spinor BECs, quantum fluctuations of atoms are expected to play a more
prominent role. Hence, their spin dynamics should be governed by spin-mixing
of quantum mechanical nature, which is responsible, for example, for the
generation of squeezed collective spin states and entangled states
\cite{Pu00}. However, the absorption-imaging technique, which works well with
large condensates, is no longer an effective detection tool for condensates
with small numbers of atoms as a result of the much reduced signals. Recently,
owing to the experimental realization of strong coupling between ultracold
atomic gases and cavity \cite{Essl07}, spinor BEC confined in a single-mode
cavity has received much theoretical attention
\cite{Liu09,FZhou08,Zhou09,Ying}. In particular, cavity transmission spectra
have been suggested as candidates for probing the quantum ground state
\cite{Liu09} or the quantum spin dynamics \cite{FZhou08} of a spinor BEC.

In all these proposals, one aims to learn the condensate dynamics from the
photons landing on the photoelectric detector. The detection is a process,
according to the Copenhagen interpretation of quantum mechanics, that projects
the system (condensate + photon) to one of the states that are eigenstates of
the observable being measured. This causes the condensate dynamics to be
disrupted in a random fashion and is the underlying physical mechanism behind
the measurement back-action. Motivated by the possibility that such a
back-action may be significant for a small condensate, we investigate, in the
present work, the effect of the back-action on the quantum spin-mixing dynamics
of a small spinor BEC subject to a homodyne detection as shown in
Fig.~\ref{scheme} (see next section for a detailed description). We take the
stochastic master equation approach which emulates the experimental process
where many runs of continuous measurements must be performed before one can
arrive at the quantum mechanical average of a dynamical observable. We show
that the effect of back-action is sensitive to whether the condensate is in the
ferromagnetic or in the antiferromagnetic ground state. We point out, in
connection with the recent interest in the dynamics of small spinor
condensates, that there is a limitation to this cavity-based detection scheme:
for sufficiently small condensates, the number of experimental runs required
to faithfully extract the internal spin dynamics can be unrealistically large;
we estimate the atom numbers above which this cavity-based homodyne detection
scheme is experimentally feasible.

The paper is organized as follows. In Sec. II we present a measurement model
and derive the stochastic master equation (SME) describing the evolution of
the spinor BEC. In Sec. III, we first analyze the measurement back-action in
two-atom case where some typical effects, such as the quantum Zeno effect
(QZE) and quantum diffusion \cite{Milburn88, Gagen92}, are clearly shown, and
then extend the analyses to $N$-atom cases by numerical simulation with
realistic experimental parameters. We show different responses of
ferromagnetic and antiferromagnetic ground state to measurements, and
measurement-induced decoherence effects in quantum spin-mixing dynamics. In
Sec. IV, the measurement outcomes are given by averaging over the
photodetector currents of repeated measurements. Finally, Sec. V concludes the paper.

\section{Model and Homodyne detection scheme}

\begin{figure}[ptbh]
\centering
\includegraphics[width=2.5in]{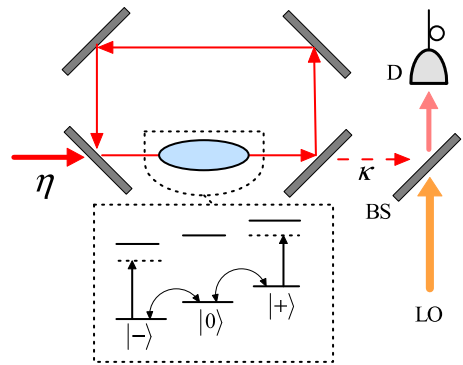} \caption{{\protect\footnotesize (color
online) Schematic diagram of homodyne detection.}}%
\label{scheme}%
\end{figure}

Figure \ref{scheme} is a schematic of our model made up of three parts: a
$F=1$ spinor BEC, a driven ring cavity, and a homodyne detection arrangement.
The BEC is assumed to be sufficiently small (with less than 1000 weakly
interacting atoms) so that\ its three spin components, $\left\vert
0\right\rangle \equiv\left\vert F=1,m_{F}=0\right\rangle $ and $\left\vert
\pm1\right\rangle $ $\equiv\left\vert F=1,m_{F}=\pm1\right\rangle $, share the
same spatial mode. In this so-called single-mode approximation (SMA)
\cite{Pu99}, we can describe the spinor condensate (subject to a quadratic
Zeeman effect) with the Hamiltonian \cite{Zhou09}%
\begin{align}
\hat{H}_{s}=  &  \hbar\lambda\lbrack(\hat{N}_{+}-\hat{N}_{-})^{2}+(2\hat
{N}_{0}-1)(\hat{N}_{+}+\hat{N}_{-})\nonumber\\
&  +2\hat{c}_{0}^{\dagger}\hat{c}_{0}^{\dagger}\hat{c}_{+}\hat{c}_{-}+2\hat
{c}_{+}^{\dagger}\hat{c}_{-}^{\dagger}\hat{c}_{0}\hat{c}_{0}]+\hbar q(\hat
{N}_{+}+\hat{N}_{-}),
\end{align}
where $\hat{c}_{i}$ is the field operator annihilating a bosonic atom in
component $\left\vert i\right\rangle $, with $\hat{N}_{i}\equiv\hat{c}%
_{i}^{\dagger}\hat{c}_{i}$ the corresponding atom number operator. $\lambda$
is a coefficient related to the spin-dependent part of the two-body
interaction, and finally $q$ the quadratic Zeeman shift.

The cavity is assumed to support a $\pi$-polarized single traveling mode with
frequency $\omega_{c}$, and is driven by an external probe field with an
amplitude $\eta$ and frequency $\omega_{p}$. As in Ref. \cite{Zhou09}, both
$\omega_{c}$ and $\omega_{p}$ are assumed to be sufficiently red-detuned from
the $F=1\leftrightarrow F^{\prime}=1$ atomic transition frequency $\omega_{a}$
so that the excited states can be adiabatically eliminated. \ Under such a
circumstance, our cavity \ + condensate system (excluding the reservoir
consisting of the cavity vacuum modes) is described by a total Hamiltonian
$\hat{H}=\hat{H}_{s}+\hat{H}_{m}$, where $\hat{H}_{m}$ is given explicitly by
\begin{equation}
\hat{H}_{m}=-\hbar\delta\hat{a}^{\dagger}\hat{a}+\hbar\eta\left(  \hat
{a}^{\dagger}+\hat{a}\right)  +\hbar U_{0}(\hat{N}-\hat{N}_{0})\hat
{a}^{\dagger}\hat{a}, \label{H_m}%
\end{equation}
with $\hat{a}$ being the field operator for annihilating a cavity photon and
$\delta=\omega_{p}-\omega_{c}$ the detuning\ of the probe relative to the
cavity mode frequency. In addition to the cavity photon energy (the first
term) and the Hamiltonian simulating the process of pumping the cavity mode by
the classical external probe field (the second term), a new term (the last
term) appears in Eq. (\ref{H_m}), which characterizes the atom-photon
interaction with an effective strength $U_{0}=g_{0}^{2}/\left(  \omega
_{p}-\omega_{a}\right)  $ with $g_{0}$ being the atom-cavity mode coupling
coefficient. \ Several comments are in order. \ First, the selection rule for
dipole transitions involving $\pi$-polarized photons only permits the
transitions between $\left\vert F=1,m_{F}=+1\text{ }\left(  -1\right)
\right\rangle $ and $\left\vert F^{\prime}=1,m_{F}^{\prime}=+1\text{ }\left(
-1\right)  \right\rangle $, and consequently, the last term, in the limit of
far-off-resonant atom-photon interaction, is expected to be proportional to
$\hat{N}_{+}+\hat{N}_{-}$ , which is equivalent to $\hat{N}-\hat{N}_{0}$ when
the definition for the total atom number, $\hat{N}=\hat{N}_{+}+\hat{N}%
_{-}+\hat{N}_{0}$, is taken into consideration. \ Second, one can
express$\ \hat{N}_{+}$ and $\hat{N}_{-}$ in terms of $\hat{N}_{0}$ along with
two constants of motion under the total Hamiltonian $\hat{H}$: $\hat{N}$ and
$\hat{M}$ $=\hat{N}_{+}-\hat{N}_{-}$ (magnetization), and as a result, we will
focus, from now on, our attention on the dynamics of $\langle\hat{N}%
_{0}\rangle$. \ Finally, we emphasize that the dispersive interaction term
$\hbar U_{0}\hat{N}_{0}\hat{a}^{\dagger}\hat{a}$ in Eq. (\ref{H_m}) can cause
the probe field to experience a phase shift proportional to $\langle\hat
{N}_{0}\rangle$, which, in the language of measurement in quantum optics,
constitutes the (matter wave) signal we aim to determine from the measurement
of the probe field. 

These discussions lend itself naturally to the final component of our model.
To begin with, we note that $\hat{N}_{0}$ does not commute with $\hat{H}$ and
cannot serve as a quantum nondemolition measurement (QND) variable
\cite{Jacobs07}, and thus, the probe field is to remain as weak as the
measurement permits in order to minimize its back-action on the signal. As the
photons leaking from the cavity are combined with those from a strong local
oscillator prior to the measurement by the photodetector, the homodyne
detection scheme illustrated in Fig.~\ref{scheme} can greatly enhance the
signal to noise ratio while at the same time allows us to directly measure the
quadrature phase amplitude and hence the $\langle\hat{N}_{0}\rangle$-dependent
probe phase shift as discussed above. In practice, in order to gain the spin
dynamics, one must monitor the phase shift continually, and perform many runs
of experiments, each of which provides a continual stream of information about
$\langle\hat{N}_{0}\rangle$, before one can average over all the runs to
construct the ensemble average $\langle\langle\hat{N}_{0}\rangle\rangle$. The
stochastic master equation approach, which combines the system-reservoir
theory with the photon counting theory \cite{Carmichael93}, is believed to be
an excellent tool to simulate such experimental processes. This is the
approach we take in the present study.

We begin with the measurement outcome, i.e., photodetector current $I$ (for a
single run), which, after subtracting the constant part due solely to the
coherent local oscillator, reads \cite{Wiseman93, Carmichael93}%
\begin{equation}
I=2\kappa\left\langle \hat{a}+\hat{a}^{\dagger}\right\rangle +\sqrt{2\kappa
}\frac{dW\left(  t\right)  }{dt}, \label{curr1}%
\end{equation}
where $\kappa$ is the cavity decay rate, and $dW(t)/dt$ represents Gaussian
white noise with $dW(t)$ an infinitesimal Wienner increment satisfying the
It{\^o} rules, $\langle\langle dW\left(  t\right)  \rangle\rangle=0$ and
$dW\left(  t\right)  ^{2}=dt$ \cite{Gardiner85}.

For the system subject to a continuous homodyne detection, its time evolution,
conditioned on a given set of measurement outcomes, is described by the
stochastic master equation (SME)%
\begin{equation}
\frac{d\rho_{c}}{dt}=-\frac{i}{\hbar}[\hat{H},\rho_{c}]+2\kappa\mathcal{L[}%
\hat{a}]\rho_{c}+\sqrt{2\kappa}\frac{dW(t)}{dt}\mathcal{H[}\hat{a}]\rho_{c},
\label{SME1}%
\end{equation}
where $\rho_{c}$ is the conditional density matrix operator for the cavity
mode + condensate system, and $\mathcal{L}$ and $\mathcal{H}$ are the
superoperators defined as%
\begin{align}
\mathcal{L[}\hat{x}]\rho &  =\hat{x}\rho\hat{x}^{\dagger}-\frac{1}{2}\hat
{x}^{\dagger}\hat{x}\rho-\frac{1}{2}\rho\hat{x}^{\dagger}\hat{x},\nonumber\\
\mathcal{H[}\hat{x}]\rho &  =\hat{x}\rho+\rho\hat{x}^{\dagger}-\text{Tr}%
\left(  \hat{x}\rho+\rho\hat{x}^{\dagger}\right)  \rho.\nonumber
\end{align}
The first term on the right-hand side of Eq.~(\ref{SME1}) represents the
unitary evolution of the system under $\hat{H}$. \ The second term describes
the decay of the cavity, originating from coarse graining over the reservoir
degrees of freedom. The last term is related to the quantum state collapse
accompanied by the detection of each photoelectron at the detector; the fact
that it shares with the current in Eq.~(\ref{curr1}) the same noise term,
$dW(t)/dt$, indicates that the evolution of $\rho_{c}$ is indeed conditioned
on the current measurement. Both the second and the last term can affect the
dynamics of the cavity field and the spinor condensate.

To clearly show the measurement back-action on the spinor BEC, we consider that
the measurement system operates in the regime where the cavity field decays at
a rate much faster than the mean-field phase shift due both to the dispersive
atom-photon coupling, and to the two-body s-wave scattering of atoms. Under
such a condition, we can approximate $\hat{a}$ around a mean value
$\alpha\equiv\left\langle \hat{a}\right\rangle \approx\eta/\kappa$ (the field
amplitude of an empty cavity) with a small fluctuation $\hat{a}^{\prime}%
$according to $\hat{a}=\alpha+\hat{a}^{\prime}$, and eliminate the modes
defined by the bosonic operator $\hat{a}^{\prime}$ adiabatically
\cite{Gagen92}. In this way, we arrive at the dimensionless SME for the
conditional density operator $\rho_{sc}=\mathrm{{Tr_{cavity}}\rho_{c}}$ of the
spinor BEC alone
\begin{equation}
\frac{d\rho_{sc}}{d\tau}=-i[\hat{H}^{\prime},\rho_{sc}]+2\xi^{2}%
\mathcal{L[}\hat{N}_{0}]\rho_{sc}+\sqrt{2}\xi\frac{dW^{\prime}}{d\tau
}\mathcal{H[}\hat{N}_{0}]\rho_{sc}, \label{new rho}%
\end{equation}
as well as the scaled photoelectric current%
\begin{equation}
I^{\prime}=2\sqrt{2}\xi\langle\hat{N}_{0}\rangle+\frac{dW^{\prime}\left(
\tau\right)  }{d\tau}, \label{current3}%
\end{equation}
where $\tau=\left\vert \lambda\right\vert t$ is the scaled time, $dW^{\prime
}(\tau)/d\tau$ the scaled white noise, and $\xi=U_{0}\eta/\sqrt{\kappa
^{3}\left\vert \lambda\right\vert }$ the measurement strength \cite{note1}. If
we were to ignore the measurement back-action, the spinor BEC would undergo a
unitary evolution under the scaled effective Hamiltonian $\hat{H}^{\prime
}=[\hat{H}_{s}+\hbar U_{0}(\hat{N}-\hat{N}_{0})\left\vert \alpha\right\vert
^{2}-\hbar\delta\left\vert \alpha\right\vert ^{2}]/\hbar\left\vert
\lambda\right\vert $. \ In what follows, in order to highlight the essential
physics, we fix the detuning to $\delta=U_{0}N$ $\ $so that $\hat{H}^{\prime}$
becomes%
\begin{align}
\hat{H}^{\prime}=  &  \frac{\lambda}{\left\vert \lambda\right\vert }[(\hat
{N}_{+}-\hat{N}_{-})^{2}+(2\hat{N}_{0}-1)(\hat{N}_{+}+\hat{N}_{-})\nonumber\\
&  +2\hat{c}_{0}^{\dagger}\hat{c}_{0}^{\dagger}\hat{c}_{+}\hat{c}_{-}+2\hat
{c}_{+}^{\dagger}\hat{c}_{-}^{\dagger}\hat{c}_{0}\hat{c}_{0}]-q^{\prime}%
\hat{N}_{0}, \label{new H'}%
\end{align}
(after removing a constant term $\hbar q\hat{N}$) where $q^{\prime}=\left(
q+U_{0}|\alpha|^{2}\right)  /\left\vert \lambda\right\vert $ is defined as a
new quadratic Zeeman shift.

The last two terms at the right-hand side of Eq.~(\ref{new rho}) represent the
measurement back-action to the spinor condensate. The first of these is
proportional to the double commutator [$\hat{N}_{0},[\hat{N}_{0},\rho_{sc}]]$,
which represents a source of decoherence in the quantum dynamics. It tends to
damp the off-diagonal elements of the density matrix under the basis of the
measured observable $\hat{N}_{0}$. It represents one form of measurement
back-action as it originates from the fact that any measurements on the cavity
mode require the use of an output coupler to couple the cavity mode to the
field modes outside the cavity. The last term in Eq.~(\ref{new rho}) can
again be traced to the measurement induced state collapse in quantum
mechanics, which is a stochastic process and hence depends on the white noise.
The dynamics obtained directly from Eq.~(\ref{new rho}) is called the
conditional dynamics, while that obtained after the ensemble average is called
the deterministic dynamics. The last term of Eq.~(\ref{new rho}) therefore
affects the conditional, but not the deterministic dynamics. Finally, in
principle, there exists another type of measurement back-action - atom heating
due to the fluctuation of the optical dipole force \cite{Stamper08}. However,
since such a fluctuation is proportional to the gradient of the cavity field
intensity,\ we anticipate the heating effect in a traveling-wave cavity to be
much weaker than what was observed in a standing-wave cavity \cite{Stamper08},
and therefore we neglect it entirely in our work here.

\section{Spin-mixing dynamics under the continuous measurements}

\subsection{Two-Atom Case}

In this section, we apply the formalism outlined in the previous section to a
two-atom \textquotedblleft toy" model, which, in principle, can be realized in
optical lattices \cite{Bloch05}, to illustrate the influence of measurement
backaction. Due to the conservation of atom number and magnetization, a spin-1
BEC with two atoms and zero magnetization is effectively a spin-1/2 system
with two basis states $\left\vert 1\right\rangle \equiv\left\vert
0,2,0\right\rangle $ and $\left\vert 2\right\rangle \equiv\left\vert
1,0,1\right\rangle $, where $|N_{+},N_{0},N_{-}\rangle$ is a Fock state with
$N_{i}$ number of atoms in spin-$i$ component. \ In this basis, $\hat
{H}^{\prime}$ (with $q^{\prime}=0$) has the following matrix representation%
\begin{equation}
\hat{H}^{\prime}=\frac{\lambda}{\left\vert \lambda\right\vert }\left(
\begin{array}
[c]{cc}%
0 & 2\sqrt{2}\\
2\sqrt{2} & -2
\end{array}
\right)  ,
\end{equation}
which has two eigenvalues, $E_{a}=2\lambda/\left\vert \lambda\right\vert $ and
$E_{b}=-4\lambda/\left\vert \lambda\right\vert $, and two corresponding
eigenstates $\left\vert a\right\rangle =\sqrt{2/3}\left\vert 1\right\rangle
+\sqrt{1/3}\left\vert 2\right\rangle $ and $\left\vert b\right\rangle
=-\sqrt{1/3}\left\vert 1\right\rangle +\sqrt{2/3}\left\vert 2\right\rangle $.
Here, $\left\vert a\right\rangle $ is the ground state when $\lambda<0$
(ferromagnetic case) and $\left\vert b\right\rangle $ is the ground state when
$\lambda>0$ $($antiferromagnetic case).

The dynamics of any atomic observable $\hat{A}$ in a particular realization
can be constructed with the help of SME (\ref{new rho}) starting from
\begin{equation}
\frac{d}{d\tau}\langle\hat{A}\rangle=\text{Tr}\left(  \frac{d\rho_{sc}}{d\tau
}\hat{A}\right)  . \label{dA/dt}%
\end{equation}
For our case here, we find that the dynamical equations for the three
Hermitian operators defined as
\begin{align}
\hat{S}_{x}  &  =\frac{1}{\sqrt{2}}(\hat{c}_{+}^{\dagger}\hat{c}_{-}^{\dagger
}\hat{c}_{0}\hat{c}_{0}+\hat{c}_{0}^{\dagger}\hat{c}_{0}^{\dagger}\hat{c}%
_{+}\hat{c}_{-}),\nonumber\\
\hat{S}_{y}  &  =\frac{i}{\sqrt{2}}(\hat{c}_{+}^{\dagger}\hat{c}_{-}^{\dagger
}\hat{c}_{0}\hat{c}_{0}-\hat{c}_{0}^{\dagger}\hat{c}_{0}^{\dagger}\hat{c}%
_{+}\hat{c}_{-}),\\
\hat{S}_{z}  &  =\hat{N}_{0}-1,\nonumber
\end{align}
are closed and can be cast into a matrix form
\begin{align}
\frac{d}{d\tau}\left(
\begin{array}
[c]{c}%
\langle\hat{S}_{x}\rangle\\
\langle\hat{S}_{y}\rangle\\
\langle\hat{S}_{z}\rangle
\end{array}
\right)   &  =\left(
\begin{array}
[c]{ccc}%
-4\xi^{2} & \mp2 & 0\\
\pm2 & -4\xi^{2} & \mp4\sqrt{2}\\
0 & \pm4\sqrt{2} & 0
\end{array}
\right)  \left(
\begin{array}
[c]{c}%
\langle\hat{S}_{x}\rangle\\
\langle\hat{S}_{y}\rangle\\
\langle\hat{S}_{z}\rangle
\end{array}
\right) \nonumber\\
&  -2\sqrt{2}\xi\frac{dW^{\prime}}{d\tau}\left(
\begin{array}
[c]{c}%
\langle\hat{S}_{x}\rangle\langle\hat{S}_{z}\rangle\\
\langle\hat{S}_{y}\rangle\langle\hat{S}_{z}\rangle\\
\langle\hat{S}_{z}\rangle^{2}-1
\end{array}
\right)  . \label{bloch}%
\end{align}
Here the upper (lower) signs are for the antiferromagnetic (ferromagnetic)
case. The terms associated with the coefficient $-4\xi^{2}$ represent the
dampings, typical of the dynamics of an open system, which destroy the
coherence and leaves the system in a mixed state composed of eigenstates of
measured observables. In the current section, we only consider the
antiferromagnetic case ($\lambda>0)$ as it does not exhibit a qualitatively
different dynamics from the ferromagnetic case.

\begin{figure}[ptbh]
\centering
\includegraphics[width=3.2in]{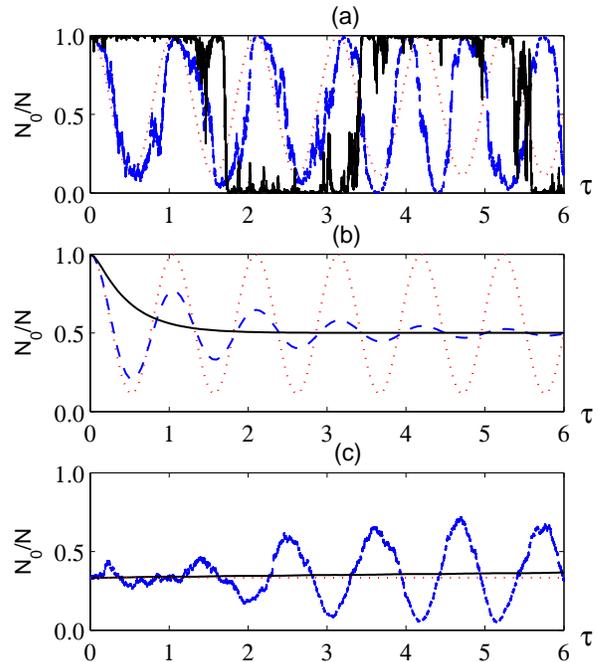} \caption{{\protect\footnotesize (color
online) The evolution of the fractional population in spin-0: $N_{0}/N$. (a)
and (b) show the conditional and deterministic evolution, respectively, with
the initial state $\left\vert 1\right\rangle $ and $q^{\prime}=0$. The red
dotted line is for non-measurement case $\xi=0$, while the blue dashed line
and black solid line is for weak measurement $(\xi=0.5)$ and strong
measurement $(\xi=2)$ case, respectively. In (c), the initial state is
$\left\vert b\right\rangle $ (see text) and $\xi=0.1$. The blue dashed line
shows the conditional evolution while the black solid line shows the
deterministic one. The red dotted line is for non-measurement case.}}%
\label{twoatoms}%
\end{figure}

Consider two atoms that are initially prepared in state $\left\vert
1\right\rangle $. Figures~\ref{twoatoms}(a) and (b) illustrate, respectively,
conditional and deterministic dynamics for systems with $\xi=0$ (red dotted
lines), $0.5$ (blue dashed lines), and $2$ (black solid lines). For the case
of no measurement ($\xi=0$), the population dynamics, $\langle\hat{N}%
_{0}\rangle/N =(1+\langle\hat{S}_{z} \rangle)/N$, undergoes a Rabi-type
oscillation with frequency $|E_{a}-E_{b}|=6$. For a relatively weak
measurement ($\xi=0.5$), besides some superimposed noises, the oscillation in
a single run begins to experience a diffusive phase shift relative to the one
without measurement; this results in a damped oscillation that one expects
when many oscillations with different phase shifts are averaged. For a
relatively strong measurement ($\xi=2$), the spin dynamics is drastically
different. This is due to that the system is ``watched" so frequently that
Quantum Zeno effect (QZE) begins to manifest itself \cite{Milburn88}. Indeed,
the conditional evolution indicates that repeated observations tend to trap
the system in states $\left\vert 1\right\rangle $ and $\left\vert
2\right\rangle $, the only two states at which the noise term in Eq.
($\ref{bloch})$ vanishes. The extra time that the system spends either in
$\left\vert 1\right\rangle $ or $\left\vert 2\right\rangle $ in the
conditional evolution slows down the transition of the initial state to other
states as shown in the deterministic evolution, which decays exponentially
without any oscillations. In both weak and strong measurements, deterministic
evolutions converge to a mixed state $(\langle\hat{S}_{x}\rangle=0,\langle
\hat{S}_{y}\rangle=0,\langle\hat{S}_{z}\rangle=0)$, the only fixed-point of
the deterministic part of Eqs. (\ref{bloch}) at which the density matrix takes
the diagonal form: $\rho_{sc}=0.5\left\vert 1\right\rangle \left\langle
1\right\vert +0.5\left\vert 2\right\rangle \left\langle 2\right\vert $.

Let us now discuss Fig. \ref{twoatoms} (c), which displays the dynamics of two
atoms initially prepared in the antiferromagnetic ground state $\left\vert
b\right\rangle $. In the absence of any measurements, as expected, the system
stays in its ground state (red dotted line). For a weak measurement ($\xi
=0.1$), the system develops a Rabi-type oscillations in the conditional
evolution, and is shown to attempt to converge to the mixed state in the
deterministic evolution (black solid line). As before, QZE appears (not shown)
when the measurement is sufficiently strong. In the two-atom case, a system
starting from the antiferromagnetic ground state exhibits similar dynamics as
that starting from the ferromagnetic ground state. However, in the $N$-atom
case, as we show in the subsection below, due to the difference in the energy
level structure and quantum fluctuation of $\hat{N}_{0}$, the measurement
back-action will have quite distinct effects on the ferromagnetic and
antiferromagnetic ground states.

\subsection{$N$-Atom Case: Conditional Population Dynamics}

Now we illustrate the measurement back-action effect for a condensate with
$N=100$ atoms. We also adopt realistic parameters: $\kappa=2\pi\times100$ MHz,
$g_{0}=2\pi\times1.6$ MHz, and $\lambda=2\pi\times20$ Hz for sodium atoms with
a typical density $10^{14}$ cm$^{-3}$ \cite{Lett09}. \ We estimate that the
value of $\xi$ lies in the range between $10^{-3}$ and $10^{-1}$ which means
measurements here are always very weak. It is impossible to find a set of
observables that are closed under Eq.~(\ref{dA/dt}) as in the two-atom case.
However, under the assumption of perfect detection (with unit detection
efficiency), we can unravel SME (\ref{new rho}) into a equivalent stochastic
Schr{\"{o}}dinger equation (SSE) \cite{Carmichael93} in the sense $\rho
_{sc}=\left\vert \psi_{sc}\right\rangle \left\langle \psi_{sc}\right\vert $
as
\begin{align}
\frac{d}{d\tau}\left\vert \psi_{sc}\right\rangle  &  =\left[  -i\hat
{H}^{\prime}-\xi^{2}(\hat{N}_{0}-\langle\hat{N}_{0}\rangle)^{2}\right.
\nonumber\\
&  \left.  +\sqrt{2}\xi(\hat{N}_{0}-\langle\hat{N}_{0}\rangle)\frac
{dW^{\prime}}{dt}\right]  \left\vert \psi_{sc}\right\rangle , \label{SSE}%
\end{align}
which allows the dynamics of any observables to be calculated exactly. The
simulation is performed using a fourth-order Runge-Kutta method for the
deterministic part, and a first-order stochastic Runge-Kutta method for the
noise part.
Furthermore, the SSE (\ref{SSE}) shows more clearly that the measurement
back-action effects are dependent not only on the measurement strength, but
also on the quantum fluctuation of the measured observable $\hat{N}_{0}$.

For the antiferromagnetic case, the ground state for $\hat{H}^{\prime}$ (with
$q^{\prime}=0$) is unique and is given by a superposition of all the Fock
states in which the spin-$1$ and $-1$ components share the same atom number:%
\begin{equation}
\left\vert \psi\right\rangle =\sum_{k=0}^{[N/2]}A_{k}\left\vert
k,N-2k,k\right\rangle ,\nonumber
\end{equation}
where the amplitudes $A_{k}$ obey the recursion relation \cite{Law98}%
\begin{equation}
A_{k}=-\sqrt{\frac{N-2k+2}{N-2k+1}}A_{k-1}.\nonumber
\end{equation}
In this state, the average atom numbers in each spin component are equal,
i.e., $\left\langle N_{+}\right\rangle =\left\langle N_{-}\right\rangle
=\left\langle N_{0}\right\rangle =N/3$, and the number-fluctuation in spin-$0$
component is super-Poissonian for $N\gg1$ \cite{Law98}. This indicates that
the antiferromagnetic ground state will be quite sensitive to the measurement
back-action effect.

\begin{figure}[ptbh]
\centering
\includegraphics[width=3.4in]{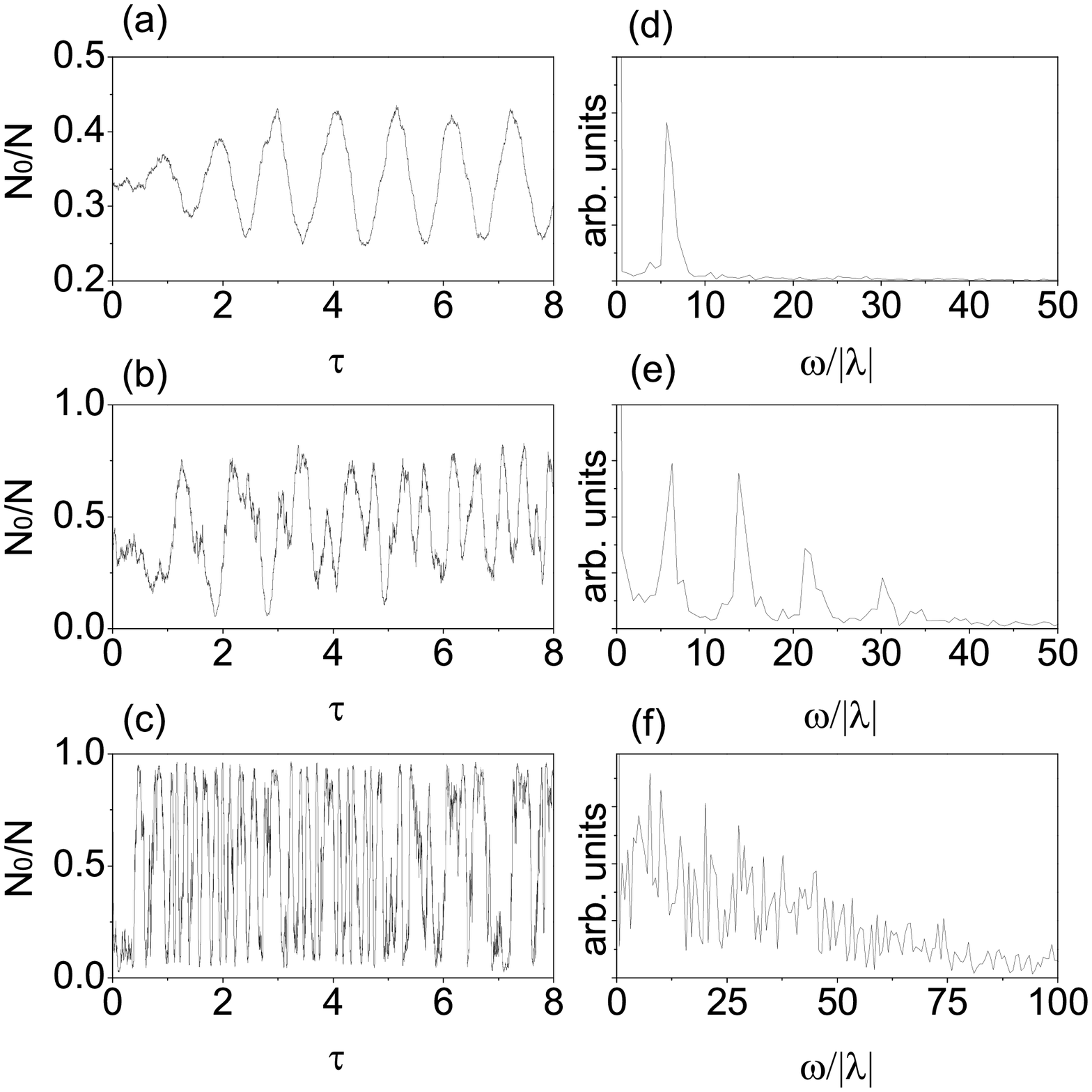} \caption{{\protect\footnotesize Left
column: Evolution of normalized particle number in spin-$0$ component under a
single run of measurements with different strength (a) $\xi=0.001$, (b)
$0.01$, and (c) $0.1$. The initial state is the anti-ferromagnetic ground
state (see the text) with total atom number $N=100$. The corresponding Fourier
spectra are shown in the right column.}}%
\label{antiferr}%
\end{figure}

Figures \ref{antiferr} illustrate the conditional population dynamics of atoms
initially prepared in the antiferromagnetic ground state and the corresponding
Fourier spectra for various measurement strengths. As Fig.~\ref{antiferr}(a)
illustrates, a measurement as weak as $\xi=0.001$ can induce the system to
oscillate predominantly at frequency $6\lambda$ (Fig.~\ref{antiferr}(d)), the
first excited frequency of the many-body system described $\hat{H}^{\prime}$
in Eq.~(\ref{new H'}). As the measurement strength increases, more and more
high-frequency components contribute to the evolutions. Fig.~\ref{antiferr}(b)
is produced with $\xi=0.01$ which is ten times stronger than in
Fig.~\ref{antiferr}(a). Indeed, instead of one peak, its Fourier spectrum
[Fig.~\ref{antiferr}(e)] displays four peaks
corresponding to the first fourth excited frequencies. Increasing $\xi$ by
another factor of ten to $\xi=0.1$ leads to a more chaotic evolution as
confirmed both by the population dynamics in Fig. \ref{antiferr}(c) and\ by
the corresponding Fourier spectrum in Fig.\ref{antiferr} (f). Here, as a
result of a dramatic increase in the number of eigenstates to which the system
can collapse, QZE becomes more complicated. In Fig. \ref{antiferr}(c), only
the transitions to $\left\vert N/2,0,N/2\right\rangle $ and $\left\vert
0,N,0\right\rangle $ are (dimly) visible because these two have the smallest
transition moments to their neighboring eigenstates. In order to see QZE
involving other transitions, we find from our numerical simulations that
strong measurements with $\xi>10$ are typically needed. 

\begin{figure}[ptbh]
\centering
\includegraphics[width=3.4in,]{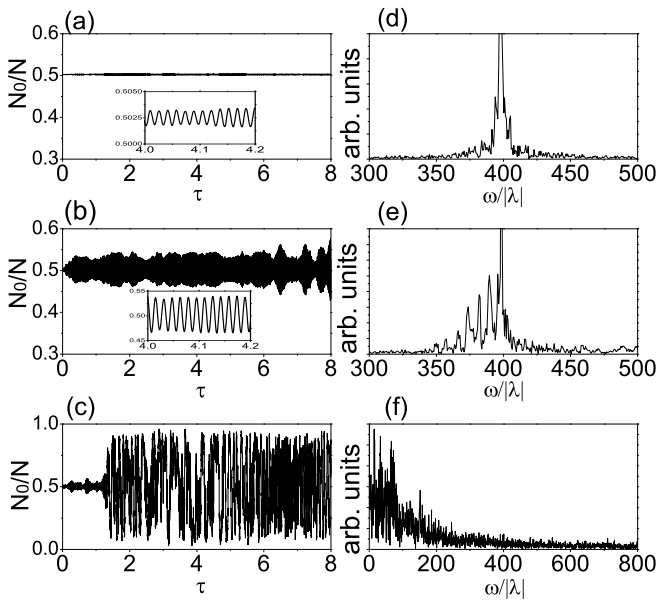} \caption{{\protect\footnotesize Same as
Fig. \ref{antiferr}, except that for a ferromagnetic condensate. Insets show
details of the oscillations. }}%
\label{ferr}%
\end{figure}

The measurement back-action on ferromagnetic spinor condensates ($\lambda<0$)
will take somewhat different effects because their distinct energy-level
structures and quantum statistical properties. The ferromagnetic ground state
is $2N+1$-fold degenerate, which reads \cite{Law98}%
\[
\left\vert \psi_{m}\right\rangle =\sum_{k}B_{k}^{(m)}\left\vert
k,N-2k-m,k+m\right\rangle ,
\]
where $m=0,\pm1,\ldots,\pm N$. Contrary to the antiferromagnetic ground state,
these states possess sub-Poissonian fluctuations in $\hat{N}_{0}$. To
demonstrate the backaction, we choose the one with $m=0$ as the initial state,
which has the largest fluctuation in $\hat{N}_{0}$.

In Fig. \ref{ferr}(a), the atomic population exhibits a weak Rabi-type
oscillations with an amplitude much smaller than that for the
antiferromagnetic case under the same measurement strength. The main reason
for this reduction is, as indicated by the Fourier spectrum in Fig.
\ref{ferr}(d), that the first excited frequency is located around $\omega
=398$, which is much higher than that for the ferromagnetic case and hence is
much difficult to excite. The reduction in the variance of $\hat{N}_{0}$ may
also weaken the effect of measurement back-action. With the increase of the
measurement strength, similar to the antiferromagnetic case, the spin
populations oscillate with multiple frequencies and become irregular with some
evidence of QZE, as shown in Figs.~\ref{ferr}(b)-(f).

\subsection{$N$-Atom Case: Comparison between Conditional and Deterministic
Population Dynamics}

The two examples considered in the previous subsection demonstrate the effect
of the measurement back-action on the population dynamics of a single
experimental realization, and as in the two-atom case, the deterministic
dynamics will emerge from the average over many runs of numerical simulations.
In order to make a smooth transition to the subject discussed in the next
section, instead of pursuing such simulations with the two examples considered
above, we seek to show the effect of averaging over many runs from a spinor
BEC initially prepared in its mean-field ground state $\left\vert \psi\left(
\tau=0\right)  \right\rangle =\left\vert 0,N,0\right\rangle $, where all the
atoms reside in the spin-$0$ component.

\begin{figure}[ptbh]
\centering
\includegraphics[width=3.2in]{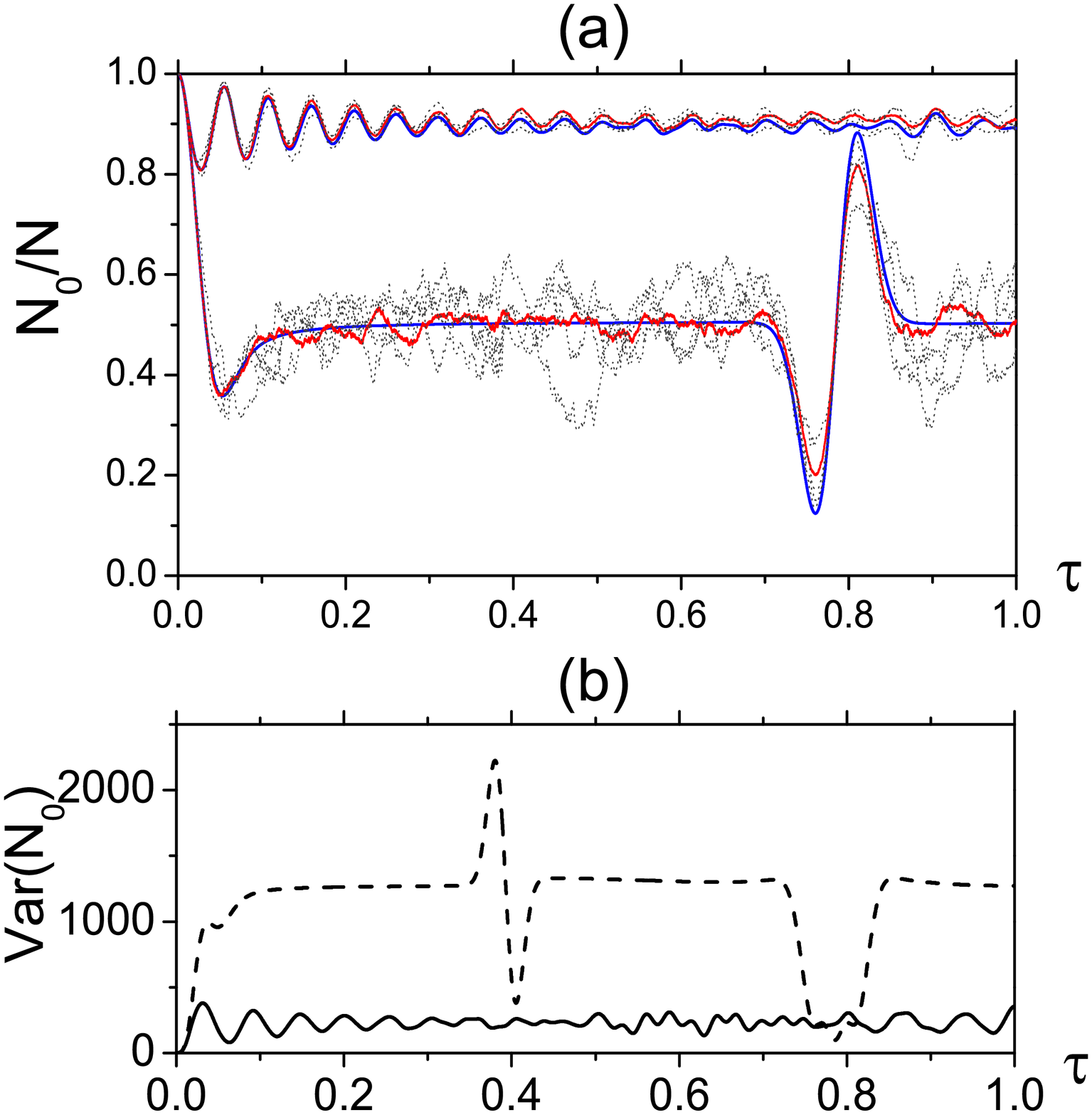} \caption{{\protect\footnotesize (color
online) (a) Time evolutions of $\langle\hat{N}_{0}\rangle/N$ for $q^{\prime
}=10$ (upper curves) and $q^{\prime}=0$ (lower curves) case with the same
initial state $\left\vert 0,N,0\right\rangle $. The blue curves correspond to
the evolution without measurement. The gray dotted curves display a variety of
evolutions conditioned on measurement outcomes with measurement strength
$\xi=0.01$, and the red curves are given by averaging over 10 conditional
evolutions. (b) The number fluctuation of spin-0 component Var($\hat{N}_{0}%
$)=$\sqrt{\langle\hat{N}_{0}^{2}\rangle-\langle\hat{N}_{0}\rangle^{2}}$, with
solid and dashed line corresponding to $q^{\prime}=10$ and $q^{\prime}=0$
case, respectively. }}%
\label{dynamics}%
\end{figure}

\begin{figure}[ptbh]
\centering
\includegraphics[width=3.2in]{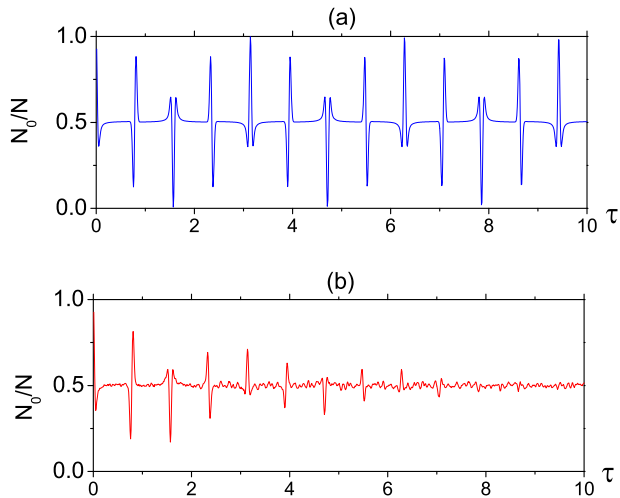}
\caption{{\protect\footnotesize (color
online) Comparing between the evolution without measurement (a)
and  the deterministic evolution (b) for the $q^{\prime
}=0$ case in a long time scale.}}%
\label{long}%
\end{figure}

The spin-mixing dynamics, in the absence of the probe, can be well understood
from the quantum-fluctuation-driven harmonic oscillator model \cite{FZhou08}.
In Fig.~\ref{dynamics}(a), we compare the spin dynamics between $q^{\prime
}=10$ (upper solid blue curve) and $q^{\prime}=0$ (low solid blue curve). In
the former case, the oscillations are weak and approximately harmonic, while
the latter case exhibits oscillations that are clearly of anharmonic nature.
This is because quantum fluctuation in $\hat{N}_{0}$ is larger for small
$q^{\prime}$ than for large $q^{\prime}$ as shown in Fig.~\ref{dynamics}(b).
The spin dynamics for $q^{\prime}=0$ in a longer time scale is shown by the
dashed blue curve in Fig.~\ref{long}(a), which clearly demonstrates a typical
quantum behavior - collapse and revival of spin oscillations. The particle
number distribution quickly collapses to a metastable regime with $\langle
\hat{N}_{+}\rangle=\langle\hat{N}_{-}\rangle=\langle\hat{N}_{0}\rangle/2=N/4$
after a time $\tau_{c}\simeq(2\sqrt{N})^{-1}$. This metastable regime is
followed by several spin oscillations and the cycle repeats itself at a time
interval $\tau_{r}=\pi$ \cite{Law98}.

In the presence of the probe, the spin-mixing dynamics will be affected by the
measurement back-action. Various conditional evolutions with an intermediate
measurement strength $\xi=0.01$ are plotted in gray dotted curves in
Fig.~\ref{dynamics}(a). For the case $q^{\prime}=10$, they are not much
different from the non-measurement evolution, so that averaging over 10
conditional evolutions appears sufficient to reveal the deterministic spin
dynamics. The quantum measurement back-action are restrained by the small
quantum fluctuation. In contrast, for the case $q^{\prime}=0$, they differ
from the non-measurement evolution quite appreciably. In this case, averaging
over 10 conditional evolutions (solid red curve) cannot produce the
anticipated deterministic dynamics and the match is particularly poor in the
metastable regime due to the large quantum fluctuation there. It requires more
runs of measurements to reveal the deterministic spin evolution. 
The curve in Fig.~\ref{long}(b) represents its deterministic evolution given by
averaging over 100 conditional evolutions, which in a short time scale traces
out the anharmonic spin oscillations clearly but indicates that the
oscillations are gradually damped for a long time evolution. The damping rate
is proportional to the measurement strength. Finally, the BEC converges to a
mixed state characterized with a diagonal density matrix, $\rho_{sc}%
=\sum_{k=0}^{N/2}P_{k}\left\vert k,N-2k,k\right\rangle \left\langle
k,N-2k,k\right\vert $, where the probability distribution function $P_{k}$ is
found (not shown) to be a constant independent of $k$ or $P_{k}=1/(N/2+1)$ to
be precise. All these are due to the decoherence induced by measurement as
discussed in the two-atom case.

\section{The Measurement Outcome: Photoelectric Current}

The numerical simulations we have considered so far show that although each
run results in a different conditional evolution $\langle\hat{N}_{0}\rangle$,
an ensemble average over dozens of these runs can already capture quite well
the deterministic quantum spin-mixing dynamics. However, what is accessible in
experiments is not $\langle\hat{N}_{0}\rangle$ but the photoelectric current
[Eq. (\ref{current3})]. Thus, in practice, $\langle\langle\hat{N}_{0}%
\rangle\rangle$ must be inferred by averaging the current over many runs of
measurements. As it turns out, it requires far more runs to reveal
$\langle\langle\hat{N}_{0}\rangle\rangle$ indirectly from the ensemble average
of the photoelectric current than directly from the ensemble average of
conditional population dynamics.

\begin{figure}[ptbh]
\centering
\includegraphics[width=3.2in]{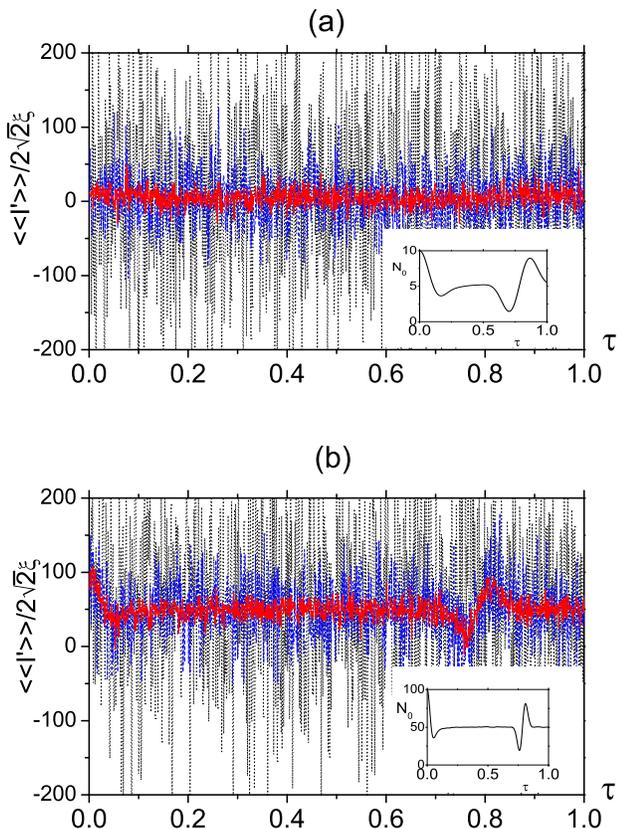} \caption{{\protect\footnotesize (color
online) Measurement outcomes given by averaging over $10^{2}$ (black),
$10^{3}$ (blue), and $10^{4}$ (red) photodetector currents. The total atom
number $N$ is $10$ in (a) and $100$ in (b), and other parameters are the same
as in Fig. \ref{dynamics}. The insets show the deterministic evolution of
$\langle\hat{N}_{0}\rangle$.}}%
\label{current}%
\end{figure}

Figures \ref{current} (a) and (b) show the ensemble averages of the
photodetector current $I^{\prime}$ for $N=10$ and $N=100$, respectively. The
black curves represent the results given by $100$ runs of measurements. As can
be seen, it is virtually impossible to extract the deterministic evolutions of
$\langle\langle\hat{N}_{0}\rangle\rangle$ (shown in the insets) as they are
dominated by white noise. In principle, one can suppress the noise by
averaging over more and more currents; this is evident from the examples
obtained when we increase the number of runs to 10$^{3}$ (blue curve) and then
to 10$^{4}$ (red curve). However, only in $N=100$ case can the spin
oscillations of deterministic nature be (vaguely) recognized. As for $N=10$
case, averaging the current over $10^{4}$ runs of measurements still do not
allow us to extract the signal.

It appears that one could increase the measurement strength $\xi$ instead of
the number of runs to enhance the signal to noise ratio according to Eq.
(\ref{current3}). However, in quantum measurements, the system dynamics is
conditioned on the detection outcomes; increasing the measurement strength
also enhances the quantum measurement back-action. First, according to the
discussion in Sec. II, strong measurement renders a large decoherence to the
measured quantum state so that the spin oscillations is rapidly damped.
Second, an increase in $\xi$ will increase the white noise in the stochastic
Schr\"{o}dinger equation, which in turn demands more runs to recover the
deterministic dynamics. Thus, there is a limitation to what we can do to
improve the signal to noise ratio by increasing the measurement strength.

An alternative is to increase the atom number. Not only does it enhance the
signal part of the current in Eq.~(\ref{current3}), but also it reduces the
quantum fluctuation of $\hat{N}_{0}$ and thus the related measurement
back-action. The net effect is the reduction in the number of required runs.
But, as the atom number increases, the mean-field dynamics will gradually
dominate \cite{Heinze10}, defeating the goal of extracting beyond-mean-field
quantum dynamics from this measurement scheme. A possible way to increase the
signal to noise ratio without raising the atom number is to process the
current signal using\ methods such as filtering high frequency components and
averaging over sliding windows \cite{Santamore04}. However, our study shows
that more than $10^{3}$ runs are still needed before we can reveal the spin
dynamics for a small BEC with less than $100$ atoms.

\section{Conclusion and Remarks}

In this work, we have considered a homodyne detection scheme, which is
designed to make a continuous measurement of the quantum spin-mixing dynamics
of a small $F$=1 spinor BEC inside an optical cavity. Using the stochastic
master equation approach, we have performed a detailed study of the quantum
measurement back-action on the spin population dynamics in the bad cavity
limit. We have used a simple two-atom system to illustrate both the
measurement-induced quantum Zeno effect and the measurement-induced diffusive
quantum dynamics. We have applied the physical intuitions gained from the
two-atom model to understand the measurement back-action on the spin population
dynamics in a spin-1 BEC. We have shown that the effect of back-action is
sensitive to the quantum fluctuation of the spinor condensate. Finally, we
stress that this study is motivated by recent proposals for using measurement
techniques popular in cavity quantum optics to probe the quantum dynamics of
small condensates. An important point we aim to make in this work is that when
applying optical detection techniques to small condensates, one needs to pay
close attention to quantum fluctuations, which are typically ignored for large
condensates. Indeed, we have shown that due to the quantum measurement
back-action, the number of runs of measurements, needed to recover the
deterministic population dynamics from the ensemble average of the
photoelectric current, increases as the number of atoms in the condensate
decreases, suggesting that the scheme is not practical for sufficiently small
condensates where the number of runs can become unrealistically large.


\section{Acknowledgments}

We thank JM Geremia for helpful discussions. This work is supported by the
National Basic Research Program of China (973 Program) under Grant No.
2011CB921604, the National Natural Science Foundation of China under Grant No.
10588402 (W.Z.), No. 11004057 (L.Z.), and No. 10874045, the Program of
Shanghai Subject Chief Scientist under Grant No. 08XD14017, Shanghai Leading
Academic Discipline Project under Grant No. B480 (W.Z.), the \textquotedblleft
Chen Guang\textquotedblright\ project supported by Shanghai Municipal
Education Commission and Shanghai Education Development Foundation (L.Z.) and
the Fundamental Research Funds for the Central Universities (L.Z., K.Z.), and
US National Science Foundation (H.P.,H.Y.L.), US Army Research Office
(H.Y.L.), and Welch Foundation with Grant No. C-1669 (H.P.).


\begin{thebibliography}{99}
\bibitem {note1}We note that $\xi$ is negative in red-detuned case but we only
give its absolute value in the following discussions.
\end{thebibliography}
\end{document}